\documentclass{article}

\usepackage{cite}

\usepackage{amsmath,amssymb,amsfonts}
\usepackage{algorithmic}
\usepackage{graphicx}
\usepackage[caption=false,font=footnotesize]{subfig}
\usepackage{textcomp}
\usepackage{balance}
\usepackage{xcolor}
\usepackage{authblk}
\usepackage{colortbl}
\usepackage{multirow}
\usepackage{array}
\usepackage[utf8]{inputenc}
\usepackage{dblfloatfix}
\usepackage{anyfontsize}
\usepackage{comment}
\usepackage{subcaption}
\usepackage[a4paper]{geometry}
\newgeometry{
    left=2.2cm,    
    right=2.2cm    
}

\usepackage{xcolor,colortbl}
\usepackage{tcolorbox}

\definecolor{Gray}{gray}{0.85}
\definecolor{gray}{rgb}{0.75, 0.75, 0.75}
\definecolor{gray_dark}{rgb}{0.70, 0.70, 0.70}
\definecolor{gray_light}{rgb}{0.86, 0.86, 0.86}

\begin{document}

\title{Grid Topology Optimization for Congestion Management Under High Renewable Penetrations and Discrete Load Growth}

\author[1]{Giacomo Bastianel}
\author[1]{Hakan Ergun}
\author[2]{Line Roald}

\affil[1]{{Department of Electrical Engineering, KU Leuven, Leuven, Belgium, and Energy Transmission Competence Hub (Etch) - EnergyVille, Genk, Belgium} \newline}
\affil[2]{{Department of Electrical Engineering, University of Wisconsin-Madison, Madison, Wisconsin, United States}}

\date{}

\maketitle

\begin{abstract}
Transmission grids are increasingly stressed by the fluctuating nature of renewable energy sources and by increasing electricity demand.
Such grids were mainly built decades ago for a fundamentally different generation fleet and load conditions. In addition, grid reinforcement projects take several years to be built and are often delayed, increasing the risk of grid congestion. Grid topology optimization offers the possibility to redistribute power flows 
by actively modifying the busbar topology of substations in the grid. 
However, the combinatorial explosion of feasible busbar configurations makes real-time topology optimization impractical for system operators. This paper proposes a methodology to identify a small subset of high-value busbar topologies to capture the economic benefit of topology optimization across a range of renewable-demand patterns. The grid topology optimization model is based on a linear-programming approximation of the AC optimal power flow formulation, and AC-feasibility checks of the optimal topology applied to the IEEE 118-bus test case. In the test case, we select two distinct pairs of substations (\textit{46--49} and \textit{24--69}) and we optimize their topology separately over 365 clustered timesteps with different wind and load conditions. For each pair, we identify the four most recurrent optimal topologies and evaluate their performance under standard, and congested conditions, with and without a discrete load growth. Results show that selecting from this reduced set of topologies reduces total generation costs by up to 0.147\% 
compared to a plain AC-OPF. Validation against a full year-long time series confirms that the identified topology subsets do reduce the total generation costs even beyond the clustered wind-load conditions. In addition, we show the influence of topology optimization on the hosting capacity of selected busbars under discrete load growth. Our findings provide a practical methodology for system operators to select a subset of optimal busbar topologies to be used in their grid for different wind-load conditions, resulting in decreasing generation costs without the computational and operational risks of real-time switching decisions.
\end{abstract}

\textit{Keywords}: Busbar Topologies Selection, Discrete Load Growth, Grid-Enhancing Technologies, Grid Topology Optimization, Renewable Energy Sources. 



\newcommand{\acnodes}{\mathcal{I}}
\newcommand{\acbranches}{\mathcal{L}}
\newcommand{\acswitches}{\mathcal{SW}^{ac}}
\newcommand{\actopology}{\mathcal{T}^{ac}}
\newcommand{\actopologyrev}{\mathcal{T}^{ac, rev}}
\newcommand{\acswitchtopology}{\mathcal{T}^{\text{sw,ac}}}
\newcommand{\acswitchtopologyrev}{\mathcal{T}^{\text{sw}^{\text{ac, rev}}}}
\newcommand{\acZILtopology}{\mathcal{T}^{\text{ZIL,ac}}}


\newcommand{\nodevoltage}{V_i}
\newcommand{\acbranchflow}{S_{lij}}



\newcommand{\dcnodes}{\mathcal{E}}
\newcommand{\dcbranches}{\mathcal{D}}
\newcommand{\dcswitches}{\mathcal{SW}^{dc}}
\newcommand{\dctopology}{\mathcal{T}^{dc}}
\newcommand{\dctopologyrev}{\mathcal{T}^{dc, rev}}
\newcommand{\dcswitchtopology}{\mathcal{T}^{\text{sw,dc}}}
\newcommand{\dcswitchtopologyrev}{\mathcal{T}^{\text{sw}^{\text{dc}, rev}}}
\newcommand{\dcZILtopology}{\mathcal{T}^{\text{ZIL,dc}}}

\newcommand{\acnodesnew}{\mathcal{I'}}
\newcommand{\acZIL}{\mathcal{S}}
\newcommand{\dcnodesnew}{\mathcal{E'}}
\newcommand{\dcZIL}{\mathcal{Q}}


\newcommand{\dcbranchflow}{P_{def}}



\newcommand{\acdcconverters}{\mathcal{C}}

\newcommand{\convertertopology}{\mathcal{T}^{\text{cv}}}

\newcommand{\generators}{ \mathcal{G}}
\newcommand{\topologies}{ \mathcal{C}}

\newcommand{\loads}{\mathcal{M}}

\newcommand{\dcgenerators}{\mathcal{G}^{\text{dc}}}

\newcommand{\dcloads}{\mathcal{M}^{\text{dc}}}

\newcommand{\genconn}{\mathcal{T}^{\text{gen}}}
\newcommand{\dcgenconn}{\mathcal{T}^{\text{gen,dc}}}

\newcommand{\acloadconn}{\mathcal{T}^{\text{load}}}
\newcommand{\dcloadconn}{\mathcal{T}^{\text{load, dc}}}


\newcommand{\genpower}{ P^g_k }
\newcommand{\acloadpower}{ S^m_k }
\newcommand{\dcloadpower}{ P^m_k }
\newcommand{\converteracpower}{ S^c_l }
\newcommand{\converterdcpower}{ P^{c, dc}_l }

\newcounter{model1} \setcounter{model1}{0}
\newcounter{model2} \setcounter{model2}{0}
\newcounter{model3} \setcounter{model3}{0}
\newcounter{model4} \setcounter{model4}{0}
\newcounter{model5} \setcounter{model5}{0}
\newcounter{model6} \setcounter{model6}{0}
\newcommand{\modelone}[1]{\noindent%
	\refstepcounter{model1}\text{(M1.\arabic{model1})}\\%
}
\newcommand{\modeltwo}[1]{\noindent%
	\refstepcounter{model2}\text{(M2.\arabic{model2})}\\%
}
\newcommand{\modelthree}[1]{\noindent%
	\refstepcounter{model3}\text{(M3.\arabic{model3})}\\%
}


\section{Introduction and motivation}
Power transmission networks were historically planned and operated under relatively predictable conditions, with largely predictable power flows from fossil-fuel-based power plants toward major demand centers. 
Under these assumptions, grid assets and substation configurations were designed to accommodate \textit{expected} power flows, and operational practices were developed around a limited set of recurring network states. The ongoing decarbonization of the power sector, and society in general, is fundamentally changing these conditions. Renewable energy sources are essential for reducing greenhouse gas (GHG) emissions; however, they introduce variability and uncertainty that system operators did not face with conventional generation. In particular, the fluctuating nature of wind and solar generation can lead to \textit{unexpected} power flow patterns that differ significantly from historical system operators' experience, causing physical congestion in the power grid~\cite{JRC_2024,Congestion_US} and stressing portions of the network that were not previously critical.

To solve grid congestion and maximize the integration of renewable energy sources (RES), system operators increasingly apply grid topology optimization as a 
remedial action, along with new grid-enhancing technologies~\cite{EPRI}, to maximize the grid transmission capacity. 
Modifying the busbar topology of selected substations through busbar splitting (BuS) and optimal transmission switching (OTS)~\cite{JAO,WATT} allows to redistribute power flows and increase the grid transmission capacity. This latter fact is particularly relevant in the context of new loads such as data centers, which typically desire to be connected to the power grid within a short period of time, but create new congestion due to their large demand.

While there is well-established literature on the OTS problem~\cite{Old_OTS,Fisher2008,Hedman_2008,Hedman2009,Hedman_2011}, 
alternative formulations~\cite{Flores,PINEDA2024110620,Coffrin_primal_dual} and approximations~\cite{Bai,OTS_QC}, there is still limited work on BuS and OTS applied to select a subset of optimal go-to topologies capable of minimizing generation costs for different renewable and demand penetrations. The literature on BuS mainly focuses on optimizing a busbar topology with~\cite{Heidarifar2014,Heidarifar2016,Morsy2022} and without~\cite{Hinneck2021} security constraints using a linearised DC-OPF formulation for a single timestep. A second-order cone relaxation~\cite{SOC} and a piecewise-linear formulation (LPAC)~\cite{LPAC} have been applied, respectively by Heidarifar~\cite{Heidarifar2021} and our previous work~\cite{Bastianel_2024} for a single timestep, too. These formulations enable the optimization model to account for reactive power and voltage magnitudes, for which gaps in control were among the factors contributing to the Iberian blackout of 28 April 2025~\cite{Black_out}, and whose effective control is expected to become increasingly important in power systems with high renewable penetration~\cite{Black_out}.

\begin{figure}[h!]
    \centering
    \includegraphics[width=0.65\linewidth]{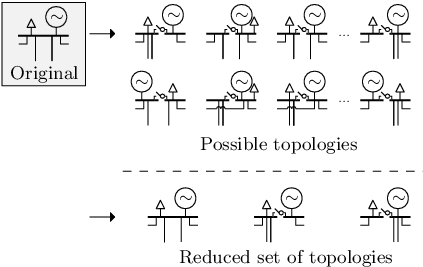}
    \caption{\small Multiple possible topologies arising from an original busbar that can be split through a busbar coupler, while the network elements connected to it can be connected to either one or the other part of the split busbars, or be disconnected from it. By selecting a reduced set of relevant topologies, the number of possible options to be selected by a system operator is decreased considerably, resuting in a reduced set of topologies.}
    \label{fig:complexity_busbar}
\end{figure}

Moreover, our previous work~\cite{Bastianel_IJEPES} has investigated day-ahead grid topology optimization accounting for RES' uncertainty in a multi-timestep and multi-scenario model. Nevertheless, all the cited references have focused on optimizing the topology of a single busbar. While topology optimization through BuS provides additional flexibility and is effective for congestion management~\cite{Sogol_congestion}, this flexibility comes with a substantial increase in operational complexity. In principle, each controllable substation can admit a large number of feasible busbar configurations, and their combinations can yield an enormous number of possible network configurations. Considering that it may be beneficial to frequently update the grid topology 
(e.g., every 15 minutes or every hour), it is impractical for a system operator to evaluate and select among thousands of candidate configurations in real time. To reduce the set of possible busbars to be split, in~\cite{Bastianel_2025_PSCC} we identified and tested metrics to select relevant busbars for grid topology optimization in a given test case. The most relevant metrics of \textit{the sum of the absolute differences in locational marginal prices (LMPs) over the branches connected to a busbar} and the presence of \textit{binding voltage angle and magnitude limits} complement the metrics identified in OTS- and BuS-related literature of the \textit{number of elements connected to the busbar} (BuS \cite{Heidarifar2021,Morsy2022}), \textit{difference in LMPs over a branch} (OTS~\cite{Ruiz_2011,Ruiz_2012}, BuS \cite{Sogol_2021,Heidarifar2021}), and \textit{congestion zones identification} (OTS~\cite{Zhou2021ACC,Khanabadi_2013}, BuS~\cite{Heidarifar2016,Heidarifar2021}). Still, when optimizing busbar configurations, generators, loads, and transmission lines can be connected to the split busbar in many different ways in our models, as shown in Figure~\ref{fig:complexity_busbar}. In addition, the congestion patterns over a grid might change over time depending on the fluctuating RES capacity factors and electrical demand levels. 

However, previous work \cite{Little_Liege} has observed that the same grid topologies tend to be optimal across a range of operating conditions. 
This suggests that it is possible to identify a reduced set of relevant topology configurations that are close to optimal, i.e. they capture most of the operational value of topology optimization in terms of generation costs reduction, but can be considered without increasing the computational cost dramatically.
To address this problem, this paper assesses whether a subset of relevant busbar configurations for selected busbars can reduce total generation costs by alleviating grid congestion, while limiting computational overhead. In addition, we analyze the effect of topology optimization of several busbar couples on load curtailment for different buses in the grid where discrete load growth takes place, and compare their hosting capacity, i.e., the additional load they can accommodate, compared to the original topology.

Therefore, this paper's main contributions are:
\begin{itemize}
    \item For a given set of busbars considered for busbar splitting, identifying a subset of busbar configurations to be used for different wind and demand conditions, leveraging the grid topology optimization model presented in~\cite{Bastianel_2024}. 
    \item Validating the identified set of busbars for a year-long time series against a plain optimal power flow model and a further reduced set of topologies.
    \item Evaluating the role of grid topology optimization in assessing a busbar's \textit{demand-growth hosting capacity} in case of a discrete load growth in selected areas in the grid. 
\end{itemize}

The paper is organized as follows. After this introduction Section, Section~\ref{sec:methodology} describes the methodology we propose to identify promising busbar configurations. Section~\ref{sec:test_case} presents the input data, i.e., the test case and renewable (wind) and demand time series. Moreover, Section~\ref{sec:results} shows the results for the selected sets of busbars in the test case. The simulations are run in \textit{standard} and \textit{congested} conditions, with and without discrete load growth. In addition, the selected busbars are validated against a year-long time series applied to the same test case. Moreover, we assess the hosting capacity for a discrete load growth for selected couples of busbars and compare the results to the original topology. Finally, Section~\ref{sec:conclusion} wraps up the paper and discusses future work.

\section{Identifying Promising Busbar Splitting Configurations} \label{sec:methodology} 

\subsection{The need for a subselection of busbar topologies}
As shown in Figure~\ref{fig:complexity_busbar}, when splitting a busbar, there is a large number of possible configurations dependent on the number of network elements connected to it. This significantly increases the complexity of finding optimal busbar splitting (BuS) configurations. Being able to limit the optimization to consider only a limited number of topologies can facilitate practical implementation for several important reasons. 

First, steady-state power flows are only one of the type of studies considered in topology optimization. Additional assessments, such as short-circuit and dynamic studies, are needed to establish the suitability of different topological options. It is therefore practical to consider only a few (pre-studied) topology options in operational optimization problems. 

Second, if the number of topologies that need to be considered is sufficiently small, it might be possible to establish approximate rules for which topology is most efficient in different kinds of operating scenarios. This could be implemented, e.g., by training a simple machine learning model on several combinations of load and RES patterns.

Third, considering only a limited number of topologies helps to significantly decrease the search space in topology optimization problems. This reduces computational overhead, providing a step towards implementing busbar splitting optimization in large systems with many busbars that can be split. 

In this paper, we seek to establish that it is possible to identify a small subset of busbar topologies that could be used as input to the development of future methods, rather than developing and testing those methods themselves. 

\subsection{Identifying a subset of promising busbar configurations}
The process to identify a subset of promising busbar topologies includes the following steps:
\begin{enumerate}
    \item Solving the topology optimization problem outlined in the next Sections for a large range of load and RES combinations.
    \item Identifying the optimal topologies as a function of load, renewable generation level, and transmission capacity. 
    \item Sub-selecting a limited number of topologies to include in future power system operations studies, based on the system operator's needs and objectives.
\end{enumerate}

\subsection{Grid topology optimization model} \label{sec:model}
\begin{figure}[h!]
    \centering
    \includegraphics[width=1.0\linewidth]{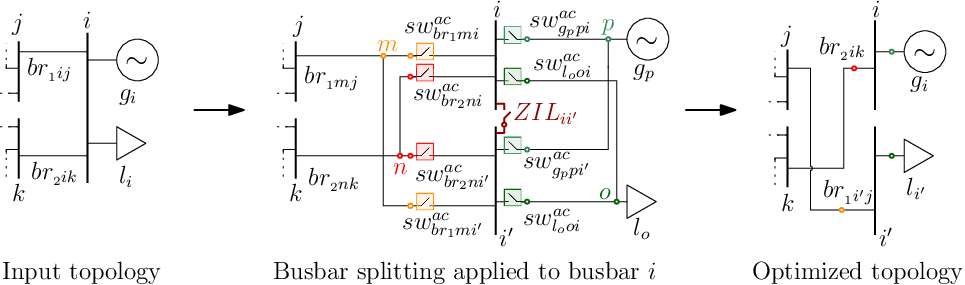}
    \caption{\small Steps behind the topology optimization model proposed in~\cite{Bastianel_2024,Bastianel_IJEPES}. Busbar $i$ (left) is split into two buses $i$ and $i'$, connected through a busbar coupler $ZIL_{ii'}$. The open/close position of the busbar coupler $ZIL_{ii'}$ is represented in the optimization model by a binary variable. Each network element that was connected to busbar $i$ in the input topology is linked to an auxiliary bus (named $m,n,o$ and $p$ in the figure) and can be connected to either one ($i$) or the other part ($i'$) of the split busbar through a switch (center). Each switch is also represented in the model through a binary variable. After the optimization, the inactive switches are removed to fix the optimized topology (right). Figure taken from~\cite{Bastianel_2025_PSCC}.}
    \label{fig:BuS_model}
\end{figure}

The grid topology optimization models we previously presented in~\cite{Bastianel_2024,Bastianel_IJEPES} find the optimal busbar topology to reduce the total generation costs by modeling BuS and optimal transmission switching (OTS) of the lines connected to the selected busbars. Therefore, each network element can be connected to either part of the split busbar, or disconnected from it. The rationale behind the proposed BuS models is represented in Figure~\ref{fig:BuS_model}.

These models rely on a linear-programming approximation (LPAC)~\cite{LPAC} of the nonconvex nonlinear AC optimal power flow (OPF) formulation. The grid topology optimization model is described by equations (M1.1)-(M1.18) in Model 1. 

\begin{table}[h!]
	\renewcommand{\arraystretch}{1.0}
	\centering
	\label{tb:Model1}
        {\fontsize{10pt}{12pt}\selectfont
	\begin{tabular}{m{40em} l}
		\hline
		Model 1: Grid topology optimization model, LPAC &\\
		\hline
		\textbf{Minimize:}\\
		 $\sum_{k \in G} c_{2k}\cdot {\genpower}^{2} + c_{1k}\cdot \genpower + c_{0k} + \sum_{ZILii' \in \acZILtopology} c_{sw}\cdot z^{sw,ac}_{ZILii'}$ & \modelone \\
         \\
         \textbf{Subject to}: & \\
	   \textbf{AC bus:}\\		
          $\theta_{r} = 0 $ \label{eq:voltage} & \modelone \\
          $(\underline{U}^{m}_{i} - 1) \leq \phi_{i} \leq (\overline{U}^{m}_{i} - 1)  \quad \forall i \in \acnodes,  $ \label{eq:voltage_magnitudes} & \modelone\\  
          $ \underline{\theta}_{i} \leq \theta_{i} \leq \overline{\theta}_{i}  \quad \forall i \in \acnodes, $ \label{eq:voltage_angles} & \modelone \\
        $\sum_{\substack{k \in \generators_i}} S^g_{k} + \sum_{\substack{l \in \loads_{i}}} S^m_{l} - Y^s_{i}(1+2\phi_{i}) + \sum_{\substack{\upsilon mi \in (\acswitchtopology_{i} \cup \acZILtopology_{i})}} {S^{sw,ac}_{\upsilon mi}} = \sum_{\substack{lij\in \actopology}} S^{ac}_{lij}$  $ \forall i \in \acnodes, $ & \modelone \\   
        \\
          \textbf{Generator}&\\
          $\underline{S}^{g}_{k} \leq S^g_k \leq  \overline{S}^{g}_{k}, \quad \forall k \in \generators \label{eq:gen_limit}$ & \modelone\\
          \\
          \textbf{AC branch}&\\
        $P_{lij} = (g_{s,ij} + g_{ij})(1 + 2\cdot\phi_{i}) -g_{ij} (cs_{ij} + \phi_{i} + \phi_{j}) - b_{ij}(\theta_{i}-\theta_{j})  \quad \forall lij\in \actopology \cup \actopologyrev$ \label{eq:Sij}  & \modelone \\
           $Q_{lij} = (b_{s,ij} + b_{ij})(1 + 2\cdot\phi_{i}) -b_{ij} (cs_{ij} + \phi_{i} + \phi_{j}) - g_{ij}(\theta_{i}-\theta_{j})  \quad \forall lij \in \actopology \cup \actopologyrev$ \label{eq:Sij}  & \modelone \\
        $|P_{lij}| \leq  \overline{P}_{lij}  \quad \forall lij \in \actopology \cup \actopologyrev\label{eq:Sijlimit}$ & \modelone\\
        $|Q_{lij}| \leq  \overline{Q}_{lij}  \quad \forall lij \in \actopology \cup \actopologyrev\label{eq:Sijlimit}$ & \modelone\\
        $cs_{ij} \leq 1 - \frac{(1 - cos(\overline{\Delta\theta_{ij}})}{(\overline{\Delta\theta_{ij}})^{2}}(\theta_{i}-\theta_{j})^{2}  \quad \forall lij \in \actopology \cup \actopologyrev$  & \modelone\\
        \\    
        \textbf{AC Busbar Splitting}&\\
        $ - (1 - z^{sw,ac}_{\upsilon mi}) \cdot M_{\theta} \leq \theta_{m} - \theta_{i} \leq (1 - z^{sw,ac}_{\upsilon mi}) \cdot M_{\theta}, \quad \forall \upsilon mi \in \acswitchtopology \cup \acZILtopology$ \label{diff_leq_M_delta} & \modelone\\
        $ - (1 - z^{sw,ac}_{\upsilon mi}) \cdot M_{m} \leq \phi_{m} - \phi_{i} \leq (1 - z^{sw,ac}_{\upsilon mi}) \cdot M_{m},  \quad \forall \upsilon mi \in \acswitchtopology \cup \acZILtopology $ & \modelone \\
        $z^{sw,ac}_{\upsilon  mi} \cdot \underline{P}^{sw,ac}_{\upsilon  mi} \leq P^{sw,ac}_{\upsilon  mi} \leq z^{sw,ac}_{\upsilon  mi} \cdot \overline{P}^{sw,ac}_{\upsilon mi},  \quad \forall \upsilon mi \in \acswitchtopology \cup \acZILtopology \label{P_ac_sw} $ & \modelone \\
        $z^{sw,ac}_{\upsilon mi} \cdot \underline{Q}^{sw,ac}_{\upsilon mi} \leq Q^{sw,ac}_{\upsilon mi} \leq z^{sw,ac}_{\upsilon mi} \cdot \overline{Q}^{sw,ac}_{\upsilon mi},  \quad \forall \upsilon mi \in \acswitchtopology \cup \acZILtopology \label{Q_ac_sw}$ & \modelone \\
        $({P^{sw,ac}_{\upsilon mi}})^2 + ({Q^{sw,ac}_{\upsilon mi}})^2 \leq z^{sw,ac}_{\upsilon mi}\cdot({\overline{S}^{sw,ac}_{\upsilon mi}})^2,  \quad \forall \upsilon mi \in \acswitchtopology \cup \acZILtopology, \label{S_ac_sw}$ & \modelone \\
        $z^{sw,ac}_{\upsilon mi} + z^{sw,ac}_{\kappa mi'} \leq 1,  \quad \forall (\upsilon mi, \kappa mi') \in \acswitchtopology  \label{z_mn_sw_ots}$ & \modelone  \\
        $z^{sw,ac}_{\kappa mi'} \leq (1 - z^{sw,ac}_{ZILii'}),  \quad \forall (\kappa mi', ZILii') \in \acswitchtopology \cup \acZILtopology \label{z_mn_sw_integer_cut}$ & \modelone  \\
        \hline
	\end{tabular}
    }
\end{table}

The objective function in (M1.1) minimizes the total generation $ \generators $ costs, where $P_{k}^{g}$ is the generation active power setpoint. A (small) switching cost $c_{sw}$ for each busbar coupler $ZILii'$ in the set of busbar couplers $\acZILtopology$ is added to split a busbar only if there is an effective economic benefit. (M1.2) sets the voltage angle of the reference bus to zero while (M1.3) and (M1.4) limit respectively the (small~\cite{LPAC}) voltage magnitude deviation $\phi_{i}$ and angles $\theta_{i}$ of the other buses between a minimum [$(\underline{U}^{m}_{i} - 1)$, $\underline{\theta_{i}}$] and a maximum [$( \overline{U}^{m}_{i}-1)$, $\overline{\theta_{i}}$] value in each node $i$ for the set of nodes $\acnodes$. Similarly, the generator setpoints are bounded by $[\underline{S}^{g}_{k}, \overline{S}^{g}_{k}]$ in (M1.6). Moreover, the AC power balance is shown by (M1.5), where $S^m_{l}$ is the nodal demand. The term $Y_{i}^{s} \cdot (1+2\phi_{i})$ refers to the power absorbed by shunt elements connected to node $i$, considered negligible in the remainder of the paper. The active ($P_{lij}$) and reactive power ($Q_{lij}$) flows through the branch $lij$ in the set of branches $\actopology \cup \actopologyrev $ are expressed in (M1.7 - M1.9). The active and reactive powers are constrained by (M1.9) and (M1.10), while 
the approximation $cs$ of the cosine $cos(\theta_{i} - \theta_{j})$ for the LPAC formulation~\cite{LPAC} is expressed in (M1.11). The switch couples ${sw_{\upsilon mi}^{ac}}$, ${sw_{\upsilon mi'}^{ac}}$ in the set $\acswitchtopology$ for each network element originally connected to the busbar and busbar couplers ${sw_{ZILii',t}^{ac}}$ in the sets of busbar couplers $\acZILtopology$ to operate splitting actions in the grid are subject to (M1.12)-(M1.18) \cite{Bastianel_2024}.
(M1.12) and (M1.13) are related to the voltage angles and the voltage magnitudes of the two buses at the extremes of each switch $sw^{ac}_{\upsilon mi}$). (M1.14), (M1.15) and (M1.16) limit the active and reactive powers of the switches to their maximum $\overline{P}^{sw,ac}_{\upsilon mi}$, $\overline{Q}^{sw,ac}_{\upsilon mi}$ and minimum values $\underline{P}^{sw,ac}_{\upsilon mi}$, $\underline{Q}^{sw,ac}_{\upsilon mi}$. (M1.17) and (M1.18) are ``exclusivity" constraints, including the switches connecting each grid element to the split busbar. The inequality constraint ($\leq$ 1) indicates that OTS of the element originally connected to the busbar and BS are both allowed in the same optimization problem, as both switches in (M1.17) are allowed to be open. Note that constraint (M1.18) imposes that if the busbar coupler is closed, i.e, BuS is not performed, the switch $sw^{ac}_{\upsilon mi,t}$ connecting the network element to the original busbar is closed. 

\begin{table}[h!]
	\renewcommand{\arraystretch}{1.0}
	\centering
	\label{tb:Model1}
        {\fontsize{10pt}{12pt}\selectfont
	\begin{tabular}{m{40em} l}
		\hline
		Model 2: Busbar topology selection among the set of optimal candidates $\topologies$ &\\
		\hline
		\textbf{Minimize:}\\
		 $\sum_{c \in \topologies}(\sum_{k \in \generators} c_{2k}\cdot {P^{g}_{k,c}}^{2} + c_{1k}\cdot P^{g}_{k,c} + c_{0k})\cdot z_{c}$ & \modeltwo \\
         \\
         \textbf{Subject to}: & \\
		\textbf{AC bus:}\\		
          $\theta_{r,c} = 0 $ \label{eq:voltage} & \modeltwo \\
          $\underline{U}^{m}_{i} \leq U^{m}_{i,c} \leq \overline{U}^{m}_{i}, \quad \forall i \in \acnodes, \forall c \in \topologies$ \label{eq:voltage_magnitudes} & \modeltwo\\  
          $ \underline{\theta}_{i} \leq \theta_{i,c} \leq \overline{\theta}_{i}, \quad \forall i \in \acnodes, \forall c \in \topologies$ \label{eq:voltage_angles} & \modeltwo \\
          $\sum_{\substack{k \in \generators_i}} S^g_{k,c} +  \sum_{\substack{lij\in \actopology}} S^{ac}_{lij,c} - y_{i,c} |U_{i,c}|^2 = 
          \sum_{\substack{l \in \loads_{i}}} S^m_l, \quad \forall i \in \acnodes, \forall c \in \topologies\label{eq:ac_power_balance}$ & \modeltwo\\
          \\
          \textbf{Generator}&\\
          $\underline{S}^{g}_{k} \leq S^g_{k,c} \leq  \overline{S}^{g}_{k}, \quad \forall k \in \generators, \forall c \in \topologies\label{eq:gen_limit}$ & \modeltwo\\
          \\
          \textbf{AC branch}&\\
        $S_{lij,c} = (Y^*_{lij,c} - \textbf{j} \frac{{b^c}_{lij,c}}{2}) \cdot  \frac{|U^{m}_{i,c}|^2}{|{T}_{lij}|^2} - Y^*_{lij,c} \cdot \frac{U^{m}_{i,c} \cdot U^{m*}_{j,c}}{{T}_{lij}}, \quad \forall lij\in \actopology, \forall c \in \topologies$ \label{eq:Sij} & \modeltwo \\
        $S_{lji,c} = (Y^*_{lij,c} -  \textbf{j} \frac{{b^c}_{lji,c}}{2}) \cdot  \frac{|U^{m}_{j,c}|^2}{|{T}_{lij}|^2} -  Y^*_{lij,c} \cdot \frac{U^{m*}_{i,c} \cdot U^{m}_{j,c}}{{T}^*_{lij}}, \quad \forall lji \in \actopology, \forall c \in \topologies\label{eq:Sji}$& \modeltwo\\
        $|S_{lij,c}| \leq  \overline{S}_{lij}, \quad \forall lij \in \actopology \cup \actopologyrev, \forall c \in \topologies \label{eq:Sijlimit}$& \modeltwo\\
        $\underline{{\theta}^{\Delta}}_{lij} \leq \angle (U^{m}_{i,c} \cdot U^{m*}_{j,c}) \leq  \overline{{\theta}^{\Delta}}_{lij}, \quad \forall lij \in \actopology, \forall c \in \topologies\label{eq:voltage_difference}$& \modeltwo\\
        \\
        \textbf{Busbar configuration}&\\
        $\sum_{c \in \topologies} z_{c} = 1 $& \modeltwo\\
        \hline
	\end{tabular}
    }
\end{table}

The optimization model selecting the (closest to) optimal busbar topology among the subset of relevant topologies selected with Model 1 is presented by equations (M2.1)-(M2.11) in Model 2. The model is comparable to a nonlinear non-convex AC-OPF model~\cite{PowerModels2018} where the OPF formulation is applied to a set of busbar topologies $\topologies$. Note that each topology $c$ consists of input data for the optimization model, and is therefore fixed. The same OPF constraints are applied to each topology $c$, but the total generation costs for the OPF of each topology would be inherently different due to the different connections of each network element within a (split) busbar. In this sense, the objective function (M2.1) minimizes the total generation costs by selecting the topology with the lowest costs through the use of the binary variable $z_{c}$. According to the special ordered set type 1 constraint (M2.11), only one binary variable, i.e., one topology, has its binary variable equal to 1. The other constraints (M2.2)-(M2.19) are the common constraints used in an AC-OPF model, with each decision variable and input parameter having an additional dimension $c$ related to each busbar topology in the subset of optimal candidates $\topologies$.

The results of the AC-OPF and LPAC-BS models are computed using a MacBook Pro with chip M1 Max and 32 GB of memory, using Gurobi v13.0~\cite{Gurobi}, MIP gap 0.05\% for the LPAC-BuS, and Ipopt~\cite{ipopt} with linear solver ma97~\cite{ma97} for the AC-OPF model.

\section{Case Study Setup}
\subsection{Test case} \label{sec:test_case}
\begin{figure}[h!]
    \centering
\includegraphics[width=0.8\linewidth]{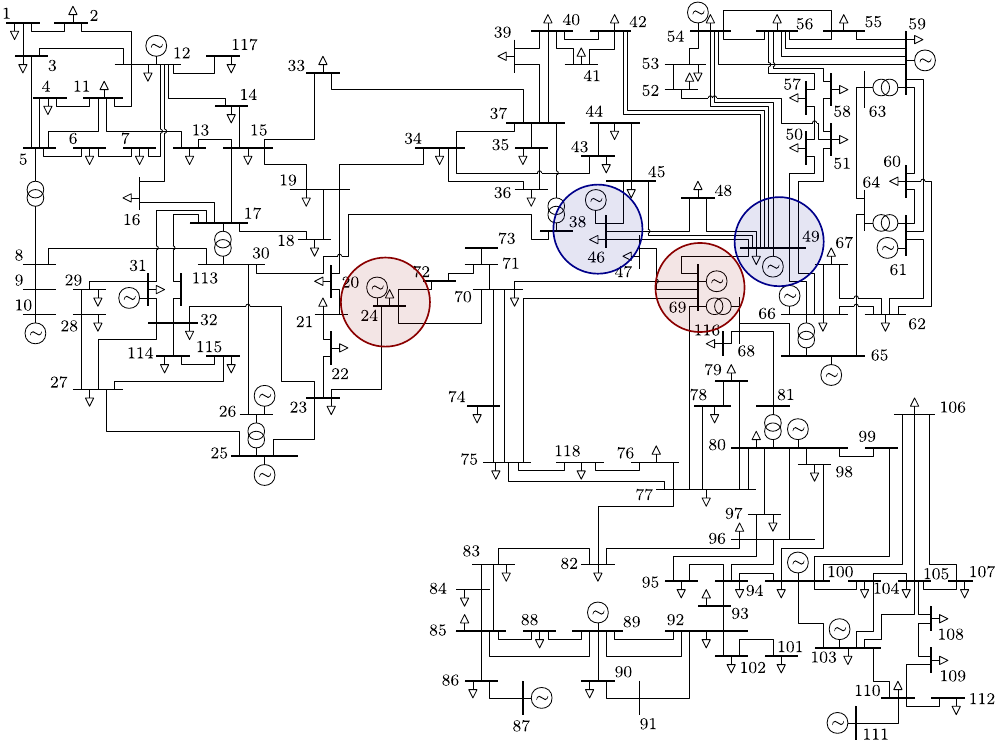}
    \caption{\small IEEE 118-bus test case~\cite{pglib} used in the paper. Grid topology optimization is applied separately to two pairs of busbars, namely \textit{24-69} (circled in red) and \textit{46-49} (circled in blue).}
    \label{fig:test_case}
\end{figure}

We apply the grid topology optimization model (Model 1) described in the previous section to the IEEE 118-bus test case~\cite{pglib} shown in Figure~\ref{fig:test_case}. The largest generator in the test case, connected to bus \textit{69}, is assumed to be a wind generator. This assumption allows us to study the congestion patterns in the test case for a changing load and generation profile, and prove how optimizing the grid topology can lead to reducing load curtailment, and therefore the total generation costs in the system. Based on the metrics developed in~\cite{Bastianel_2025_PSCC} and the results from AC-OPF simulations, we select two pairs of busbars, \textit{46-49} and \textit{24-69}, to be potentially split, as shown in Figure~\ref{fig:test_case}. These combinations of busbars lead to a reduction in generation costs of respectively 0.53\% and 0.52\% when tested in the default conditions of the IEEE 118-bus test case~\cite{pglib}.

\subsection{Renewable and demand time series} \label{sec:time_series}
\begin{figure}[h!]
    \centering
\includegraphics[width=0.75\linewidth]{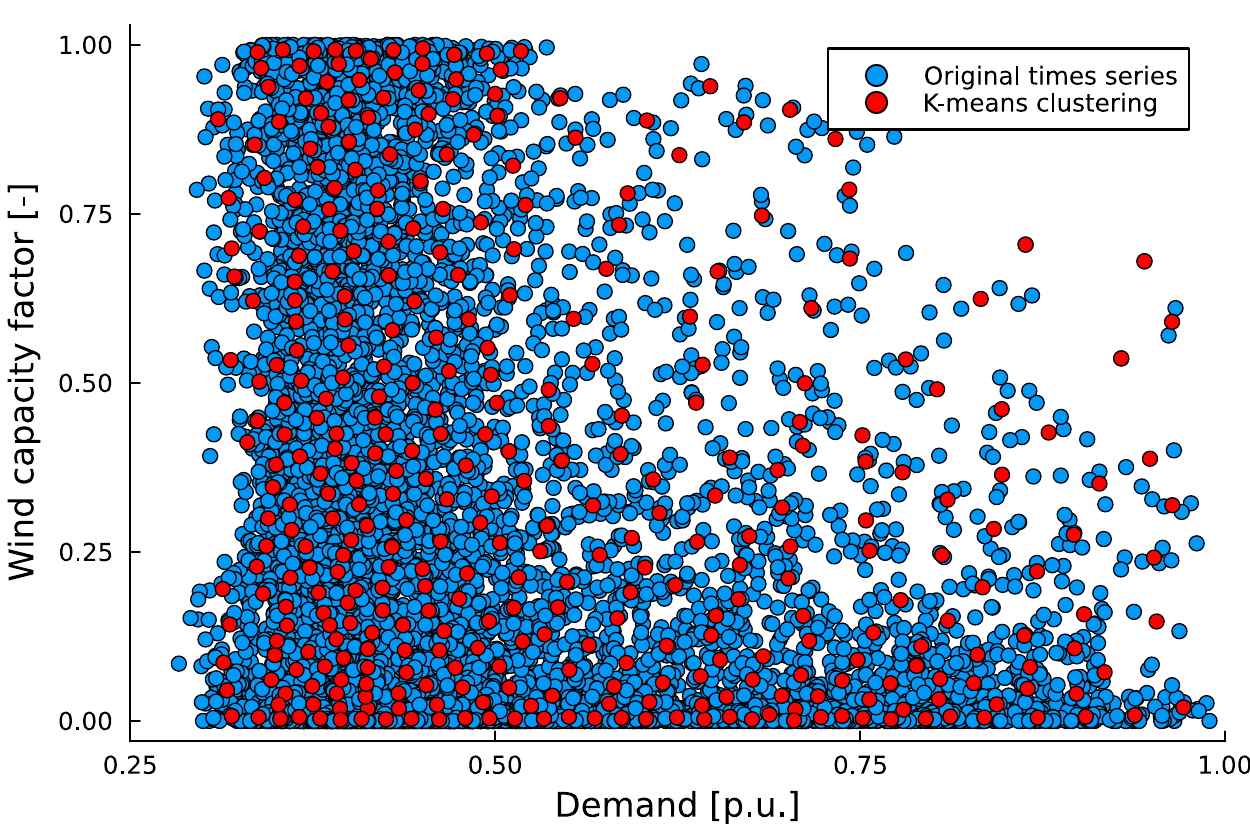}
    \caption{\small Wind and demand time series from wind generator 2 and demand center 2 from the RTS-GMLC test case~\cite{RTS_GMLC} (\textit{Original time series}), and K-means clustering~\cite{Jin2010} to reduce the number of timesteps to 365 (\textit{K-means clustering}).}
    \label{fig:time_series}
\end{figure}

The wind and demand time series used in this paper are inspired by the RTS-GMLC test case~\cite{RTS_GMLC}, day-ahead measure, wind 2, and load 2. To keep the busbar topologies selection problem tractable, we apply K-means clustering~\cite{Jin2010} to the two time series to select 365 representative time steps. Both the original time series from~\cite{RTS_GMLC} and the K-means clustering are shown in Figure~\ref{fig:time_series}.

To test the grid topology optimization models in different loading conditions, we evaluate the test case in a \textit{congested} and a \textit{standard} case:
\begin{itemize}
    \item The \textit{congested} conditions are computed by doubling each load from the original IEEE 118-bus test case, and multiplying it by the demand in each timestep from the K-means clustering. 
    \item In the \textit{standard} conditions, each load from the original IEEE 118-bus test case is only multiplied by the demand in each timestep from the K-means clustering. 
\end{itemize}
In addition, discrete load growth conditions are introduced by adding a constant load equal to the biggest load in the original IEEE 118-bus test case to bus \textit{69}. This case can represent the connection of, e.g., a large data center with a flat load profile to the test case.

\subsection{Simulations} \label{sec:simulations}
We use the grid topology optimization model presented in Section~\ref{sec:methodology} to individually optimize the topology of the two pairs of busbars introduced in Section~\ref{sec:test_case}.

Note that in this paper we assume that each \textit{substation} has one single \textit{busbar}, as is the case for the IEEE 118-bus test case. In a real-life setting, \textit{one substation} often has \textit{multiple busbars}. The proposed optimization models can still be applied in the presence of multiple busbars, but the inherent complexity of the optimization problem would increase because of the increase in the possible combinations of busbar topologies described in Section~\ref{sec:methodology}.

The model is run for each of the 365 timesteps shown in Figure~\ref{fig:time_series} in the \textit{standard} and \textit{congested} conditions mentioned in the previous Section~\ref{sec:time_series}, with and without discrete load growth. We select the four most recurrent optimal topologies for each simulation (\textit{46-49 congested}, \textit{46-49 standard}, \textit{46-49 congested load growth}, \textit{46-49 standard load growth}, \textit{24-69 congested}, \textit{24-69 standard}, \textit{24-69 congested load growth} and \textit{24-69 standard load growth}), and assign the one leading to the lowest generation costs to each of the 365 timesteps shown in Figure~\ref{fig:time_series}. These topologies are referred to as ``\textit{selected topologies}'' in the remainder of the paper.

Note that if the lowest generation costs among the \textit{selected topologies} are higher than the ones from the AC-OPF simulation of the original topology, the original topology is assigned to the timestep. 
As a result, we map subsets of timesteps in which the different \textit{selected topologies} lead to the lowest generation costs, and discuss them in Section~\ref{sec:results}. Furthermore, we show that the \textit{selected topologies} lead to lower generation costs than the OPF simulations for the original 8784 timesteps of the RTS-GMLC case. By using topologies from the subset of \textit{selected topologies}, the computational time is considerably reduced compared to finding the optimal topology for every timestep, too.

Moreover, we investigate the bus hosting capacity with a discrete load growth in some of the buses with the lowest LMPs (\textit{25}, \textit{61}, and \textit{100}) in different zones of the test case. For these simulations, these buses' \textit{demand-growth hosting capacity} is computed for a selected timestep (\textit{65}) with low demand and low wind conditions to test the system in congested conditions. We compare the \textit{demand-growth hosting capacity} of each busbar in terms of the number of times they can host an increase equal to the biggest load in the test case, 2.77 pu at bus \textit{116}. 

The \textit{demand-growth hosting capacities} being compared are those of the original topology, of the cheapest topology out of the \textit{selected topologies} for busbars \textit{46-49} and \textit{24-69} at timestep \textit{65}, and of the optimal topologies computed applying Model 1 separately to the busbar couples \textit{46-49}, \textit{24-69}, \textit{24-49}, \textit{25-66} and \textit{77-80}. The last three couples are additional busbar couples selected through the metrics presented in~\cite{Bastianel_2025_PSCC} (\textit{sum of the absolute differences in LMPs over the branches connected to a busbar} and \textit{number of elements connected to the busbar}) at the point of load shedding in the original topology for a discrete load growth in buses \textit{25}, \textit{61}, and \textit{100}. These busbar couples are selected to evaluate the developed models with couples other than the ones used previously in the paper, and selected at the point of load shedding in the system, i.e. in stressed conditions for the system.

\begin{figure}[h!]
    \centering
\includegraphics[width=0.8\linewidth]{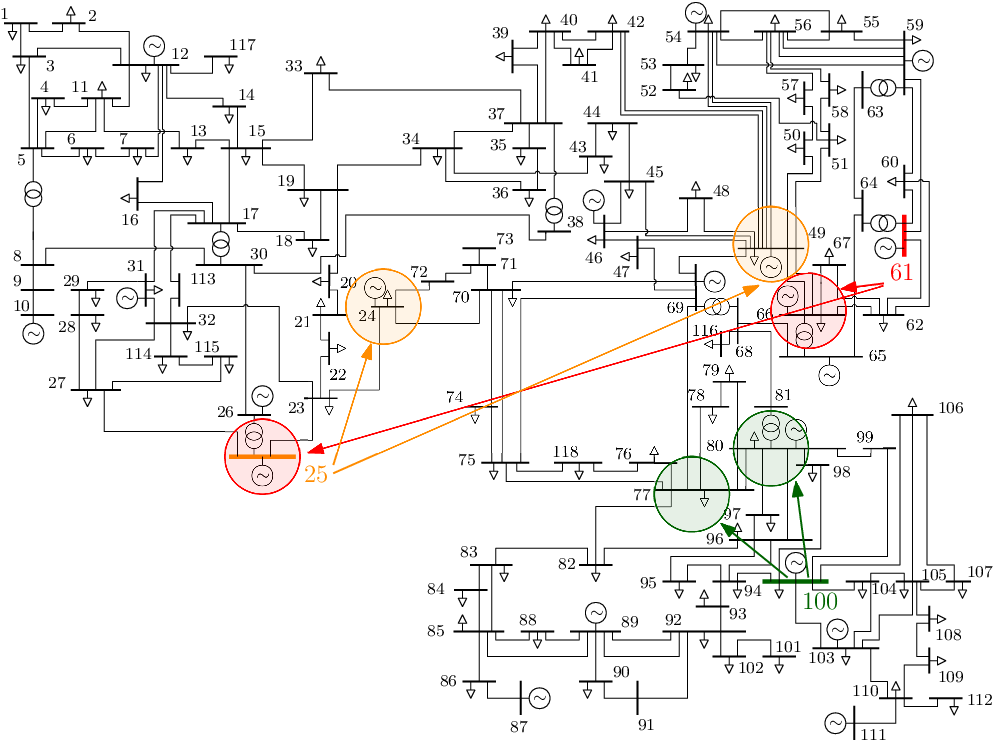}
    \caption{\small IEEE 118-bus test case~\cite{pglib} used in the paper. 
    A discrete load growth is simulated in buses \textit{25} (orange), \textit{61} (red), and \textit{100} (green) to compute the value of grid topology optimization in increasing each busbar \textit{demand-growth hosting capacity}. Using the busbar selection metrics developed in~\cite{Bastianel_2025_PSCC} to find relevant areas in the grid for grid topology optimization, couples of busbars to possibly increase each bus' \textit{demand-growth hosting capacity} through topology optimization are assigned to each bus and circled in its respective color.}
    \label{fig:test_case_load_growth}
\end{figure}

In this analysis, each low-LMP bus has a preferential ``assigned'' busbar couple (\textit{25} $\rightarrow$ \textit{24-49}, \textit{61} $\rightarrow$ \textit{25-66}, \textit{100} $\rightarrow$ \textit{77-80}) computed with the metrics from~\cite{Bastianel_2025_PSCC} mentioned before, as represented in Figure~\ref{fig:test_case_load_growth}. Nonetheless, we also run the grid topology optimization models separately for each of the ``non-assigned'' combinations, i.e. \textit{25} $\rightarrow$ \textit{25-66} and \textit{77-80}, \textit{61} $\rightarrow$ \textit{24-49} and \textit{77-80}, \textit{100} $\rightarrow$ \textit{25-66} and \textit{77-80}, to have a more complete view of the value of topology optimization for a bus \textit{demand-growth hosting capacity}.

Therefore, the final goal is to evaluate whether topology optimization can help increase the bus \textit{demand-growth hosting capacity}, and by how much.

\section{Numerical Results} \label{sec:results}
In this Section, we present the results of the simulations described previously in Section~\ref{sec:simulations}. The goal of our analysis is to show whether we can reduce the total generation costs and demand curtailment in the system by selecting the cheapest topology for two pairs of busbars (\textit{46-49} and \textit{24-69} from the IEEE 118-bus test case~\cite{pglib}) for each timestep from a subset of \textit{selected topologies}. This subset counts up to four \textit{selected topologies} identified with the topology optimization models described in Section~\ref{sec:model}. We select them by optimizing the busbar topologies for each of the 365 timesteps shown in Fig.~\ref{fig:time_series} for both the previously introduced \textit{congested} and \textit{standard} conditions, and identifying the most recurrent ones.

\subsection{Splitting busbars 46 and 49} \label{sec:opt_top_46_49}
\subsubsection{Congested conditions}
The \textit{selected topologies} for busbars \textit{46-49} in \textit{congested} conditions are shown in Fig.~\ref{fig:46_49_congested}. In three of the four \textit{selected topologies} (\textit{topology 1}, \textit{topology 3}, and \textit{topology 4}), both busbars are split, while in \textit{topology 2} only busbar \textit{49} is split, and OTS is applied to a branch connected to busbar \textit{46}. We note that most of the topological actions in these topologies are related to rearranging the connections of the branches linked to busbar \textit{49}, i.e. redistributing the power flows in this specific part of the test case.
\begin{figure}[ht]
    \centering
\includegraphics[width=0.7\linewidth]{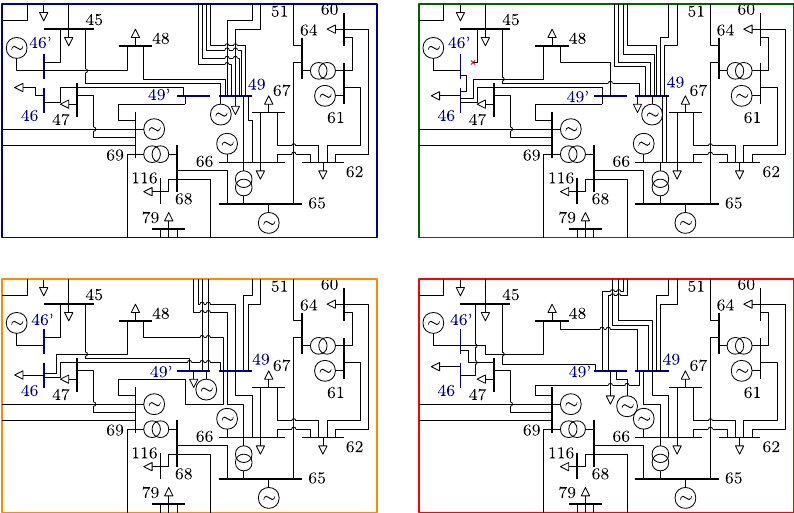}
    \caption{\small Subset of the \textit{selected topologies} for busbars \textit{46}-\textit{49}, obtained as a result of the topology optimization model presented in~\cite{Bastianel_2024,Bastianel_IJEPES} for different wind and demand conditions. In the text, they are referred to as \textit{topology 1} (top-left, blue), \textit{topology 2} (top-right, green), \textit{topology 3} (bottom-left, orange) and \textit{topology 4} (bottom-right, red).}
\label{fig:46_49_congested}
\end{figure}

Subsequently, by running the original AC-OPF formulation~\cite{PowerModels2018} for each of the four topologies and the original test case, we identify the topology with the lowest generation costs for each of the 365 timesteps shown in the previous Fig.~\ref{fig:time_series}. The distribution of such topologies is displayed in Fig.~\ref{fig:46_49_congested}. We observe that there are distinct zones in which each topology shows lower total generation costs, such as \textit{topology 1} for high demand and low wind, and \textit{topology 3} for low demand conditions. In these \textit{congested} conditions, the original grid topology (\textit{Original} in Fig.~\ref{fig:46_49_congested}) has lower costs than the \textit{selected topologies} for only 26 out of the 365 timesteps, as indicated in Table~\ref{tab:results_optimization}. Furthermore, the results in Table~\ref{tab:results_optimization} suggest that adjusting the grid topology following the identified subset of \textit{selected topologies} reduces the total generation costs (0.101\%) and load curtailment (0.295\%) compared to the OPF solutions for this \textit{congested} case.

\begin{figure}[ht]
    \centering
\includegraphics[width=0.7\linewidth]{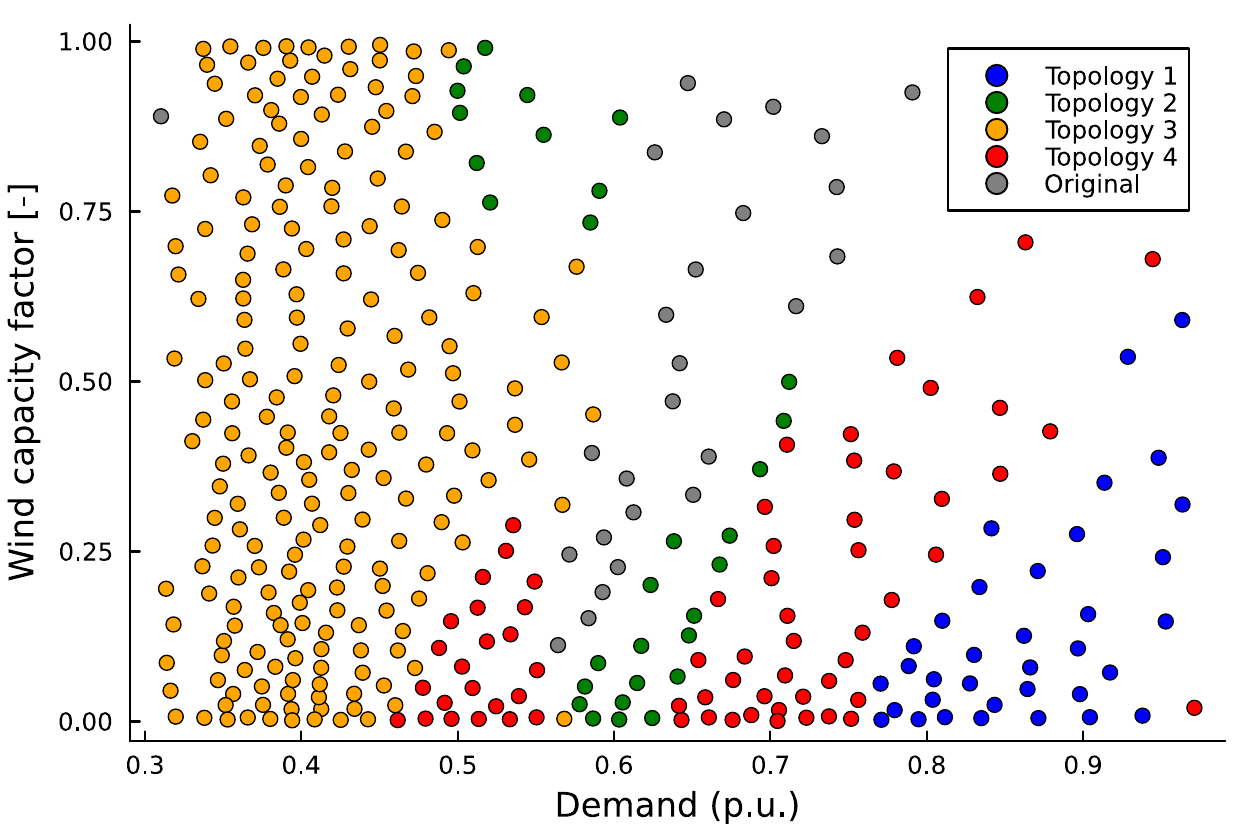}
    \caption{\small Optimal busbars \textit{46-49} \textit{selected topologies} leading to the lowest generation costs in an AC-OPF simulation for each demand-wind combination in \textit{congested} conditions. If the original test case topology has the lowest generation costs, the timestep is indicated in gray (\textit{Original}). Combinations of demand-wind in which different topologies are the most beneficial from a total costs perspective are indicated by the different color distributions along the graph.}
    \label{fig:46_49_congested}
\end{figure}

\begin{table*}[h!]
\caption{\small The topology of busbars \textit{46-49} and \textit{24-69} is optimized for four cases, namely \textit{Congested}, \textit{Standard}, \textit{Congested load growth (LG)}, and \textit{Standard load growth (LG)}, over a set of 365 timesteps. Four \textit{selected topologies} ($C_1$ $\rightarrow$ \textit{topology 1}, $C_2$ $\rightarrow$ \textit{topology 2}, $C_3$ $\rightarrow$ \textit{topology 3}, and $C_4$ $\rightarrow$ \textit{topology 4}) are identified as the most recurrent over 365 timesteps. The number of timesteps ``\# Timesteps'' indicates in how many timesteps each \textit{selected topology} (including the original one, OPF) leads to the lowest generation costs when running an AC-OPF. The ``Total generation costs'' and ``Curtailment'' are computed for each case for the OPF and topology optimization (TO) models. Their difference is expressed in terms of percentage ``Reduction''.}
\begin{center}
\resizebox{\textwidth}{!}{%
\begin{tabular}{|c|c|c|c|c|c|c|c|c|c|c|c|c|c|c|c|c|}
\hline
\multirow{2}{*}{Split} & \multirow{3}{*}{Case} & \multicolumn{5}{c|}{\# Timesteps [h]} & \multicolumn{3}{c|}{Total generation costs $\cdot10^{5}$ [\$]} & \multicolumn{3}{c|}{Curtailment [\% of total load]} \\
\cline{3-13} 
 \multirow{2}{*}{Buses}& &  \multirow{2}{*}{C$_1$} & 
 \multirow{2}{*}{C$_2$} & 
 \multirow{2}{*}{C$_3$} & 
 \multirow{2}{*}{C$_4$} & 
 \multirow{2}{*}{OPF} &
 \multirow{2}{*}{OPF} &
 \multirow{2}{*}{TO} &
 Reduction &
 \multirow{2}{*}{OPF} & 
 \multirow{2}{*}{TO} & Reduction \\
 & & & & & & & & & [\%] & & & [\%] \\
\hline
\multirow{4}{*}{46-49} & Congested    & 35 & 30 & 206 & 68 & 26 & 486.540 & 485.997 & 0.111 & 7.454 & 7.432 & 0.295\\
 & Standard  & 12 & 88 & - & - & 265 & 168.134 & 168.103 & 0.018 & 0 & 0 & - \\
 & Congested, LG & 99 & 3 & 26 & 158 & 79 & 518.333 & 517.674 & 0.127 & 8.168 & 8.144 & 0.294 \\
 & Standard, LG & 14 & 4 & - & - & 347 & 185.638
 & 185.623 & 0.008 & 0 & 0 & - \\
\hline
 \multirow{4}{*}{24-69} & Congested    & 136 & 10 & 110 & 69 & 40 & 486.540 & 486.146 & 0.091 & 7.144 & 7.138 & 0.084\\
 & Standard & 305 & 53 & - & - & 7 & 168.134 & 168.005 & 0.077 & 0 & 0 & - \\
 & Congested, LG & 60 & 83 & 78 & 12 & 132 & 518.333 & 517.674 & 0.127 & 8.168 & 8.148 & 0.245 \\
 & Standard, LG & 1 & 2 & - & - & 362 & 177.612 & 177.603 & 0.005 & 0 & 0 & - \\
\hline
\end{tabular}}
\label{tab:results_optimization}
\end{center}
\end{table*}

\subsubsection{Standard conditions}
\begin{figure}
    \centering
\includegraphics[width=0.7\linewidth]{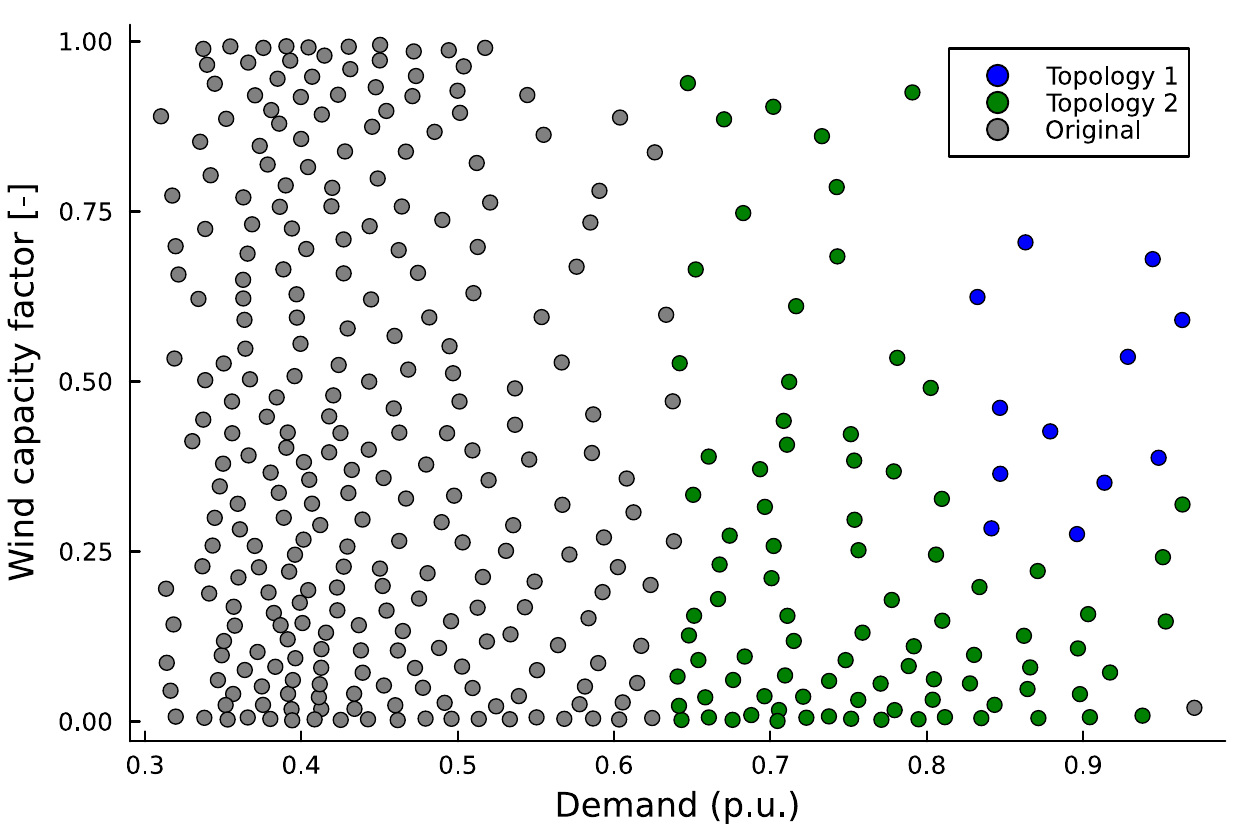}
\caption{\small Optimal busbar \textit{46-49} \textit{selected topologies} having the lowest costs in an AC-OPF simulation for each demand-wind combination in \textit{standard} conditions. With low demand levels, the \textit{original} topology dominates the distribution. With increasing congestion in the grid, the optimal busbar topologies lead to lower generation costs.}
\label{fig:46_49_standard}
\end{figure}

In case of \textit{standard} conditions, grid congestion becomes less severe in the test case compared to the \textit{congested} conditions. As a result, optimizing the busbar topology leads to different optimal topologies and lower economic benefits compared to the \textit{congested} case, as shown in Fig.~\ref{fig:46_49_standard} and Table~\ref{tab:results_optimization}. Particularly, the \textit{original} topology is the cheapest one for 265 timesteps, especially with low demand, as shown in Fig.~\ref{fig:46_49_standard}. With increasing demand, topologies \textit{1} and \textit{2} have the lowest generation costs for 12 and 88 timesteps, respectively. As a result, the cost reduction is limited to 0.016\%, with no curtailment happening in the system.

\subsubsection{Load growth at bus 69}
We simulate a discrete load growth in the system by applying to bus \textit{69} a constant demand of the same magnitude as the biggest load in the original 118-bus test case, connected to bus \textit{116} (4.34\% of the total load). This setup allows us to study whether grid topology optimization to manage congestion and reduce the total generation costs can be performed even in more congested conditions in the test case. Even though with different optimal subsets of \textit{selected topologies}, we observe a similar pattern as in the previous Fig.~\ref{fig:46_49_congested} and Fig.~\ref{fig:46_49_standard} for the \textit{congested load growth} and \textit{standard load growth} cases. To keep the text concise, we do not add figures for these two \textit{load growth} cases, but based on Table~\ref{tab:results_optimization} we notice that the results are comparable to the findings of previous Fig.~\ref{fig:46_49_congested}, i.e. that, in congested conditions, using a subset of optimal topologies does lead to lower total generation costs and reduced curtailment compared to plain AC-OPF simulations. The reduction in total costs (more than 0.1\%) and curtailment (0.295\%) is comparable to the \textit{congested} case. In the \textit{standard load growth} case, the best topology for 173 timesteps is \textit{topology 2}, followed by the \textit{original one} (\textit{OPF}). These results confirm the fact that the value of topology optimization increases with congested conditions in the grid, in this case caused by the additional load added to bus \textit{69}.

\subsection{Splitting busbars 24 and 69} \label{sec:opt_top_24_69}
\begin{figure}[h!]
    \centering
\includegraphics[width=0.7\linewidth]{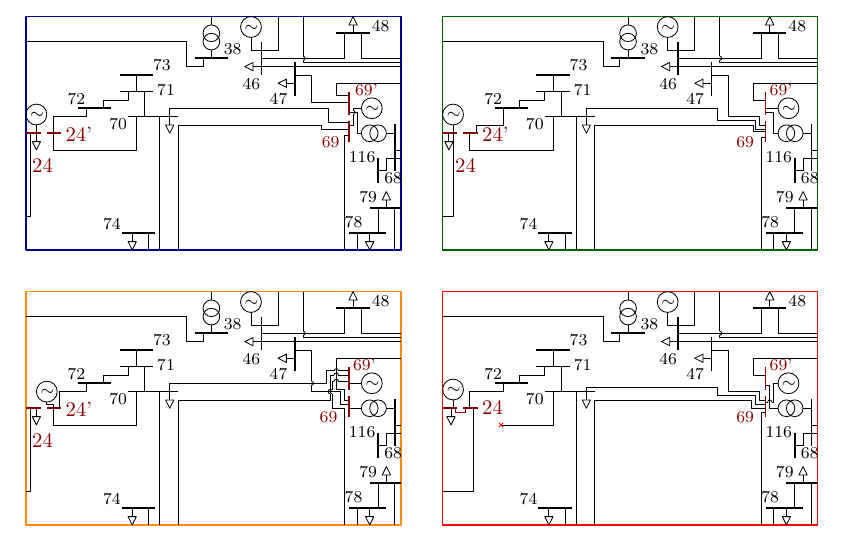}
\caption{\small Subset of the \textit{selected topologies} for busbars \textit{24}-\textit{69}, obtained as a result of the topology optimization model presented in~\cite{Bastianel_2024,Bastianel_IJEPES} for different wind-demand conditions. In the text, they are referred to as \textit{topology 1} (top-left, blue), \textit{topology 2} (top-right, green), \textit{topology 3} (bottom-left, orange) and \textit{topology 4} (bottom-right, red).}
    \label{fig:topologies_24_69}
\end{figure}

We run the same analysis as the previous section for busbars \textit{24-69} to validate the findings of our proposed methodology. Fig.~\ref{fig:topologies_24_69} shows the \textit{selected topologies} for the \textit{congested} case.

\begin{figure}[h!]
    \centering
\includegraphics[width=0.7\linewidth]{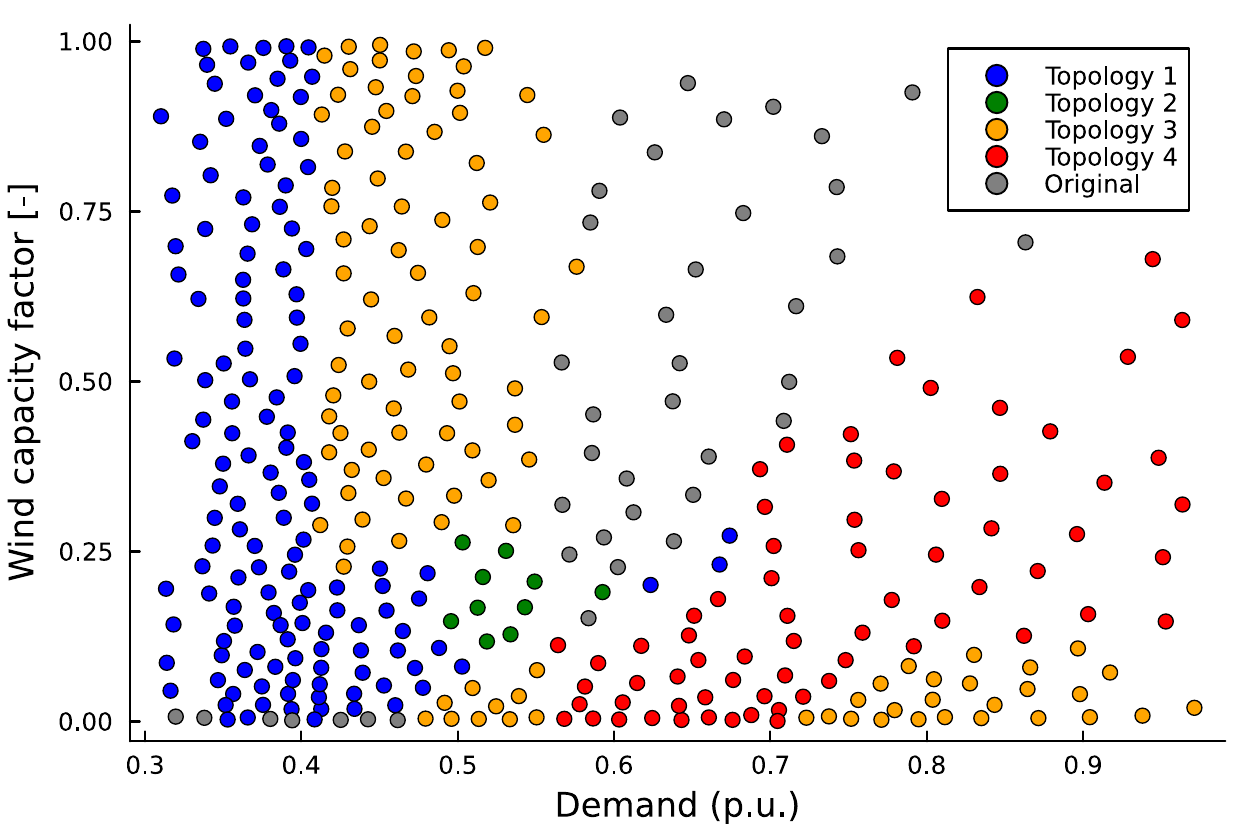}
    \caption{\small Optimal busbars \textit{24-69} \textit{selected topologies} leading to the lowest generation costs in an AC-OPF simulation for each demand-wind combination in \textit{congested} conditions. If the original test case topology has the lowest generation costs, the timestep is indicated in gray (\textit{original}). Combinations of demand-wind in which different topologies are the most beneficial from a total costs perspective are indicated by the different color distributions along the graph.}
\label{fig:69_24_congested}
\end{figure}

The \textit{congested} (Fig.~\ref{fig:69_24_congested}) and \textit{congested load growth} results for this case show a distinction in the optimal topology for different wind-demand ratios, similarly to the same cases for busbars \textit{46-49}. Moreover, the generation costs (0.091\% and 0.127\%) and the load growth curtailment (0.245\%) reductions for these two cases indicated in Table~\ref{tab:results_optimization} are comparable to the ones seen for busbars \textit{46-49}. These results for \textit{congested} conditions confirm that using a subset of busbar topologies for different wind-demand ratios leads to lower generation costs and curtailment reduction compared to the OPF results. In addition, using a subset of \textit{selected topologies} does not require the use of real-time topology optimization, which can be a source of uncertainty for system operators. 

In \textit{standard} conditions, Table~\ref{tab:results_optimization} displays how optimizing the grid topology leads to lower costs compared to the original topology for the large majority of the timesteps in the \textit{24-69} \textit{standard} case. In the \textit{standard load growth} case, though, the original topology is the cheapest for 362 timesteps. These results are opposite compared to the trends seen in the \textit{standard} cases from busbars \textit{46-49}. This difference is motivated by the fact that busbar \textit{69} has binding voltage angle limits even for the \textit{standard} conditions, and therefore benefits from the proposed topology optimization model. When a large load, i.e. physical congestion, is added to it, there is no room for additional transmission capacity obtained through topological actions. Therefore, their effect on reducing the generation costs compared to an AC-OPF is limited.

\subsection{Validation with a year-long time series}
In this part of the Section, we validate the \textit{selected topologies} from Sections~\ref{sec:opt_top_46_49} and~\ref{sec:opt_top_24_69} with a year-long time series inspired by the RTS-GMLC test case, day-ahead measure, wind 2, and load 2~\cite{RTS_GMLC}. We aim to show that using a subset of four \textit{selected topologies} leads to a significant decrease in generation costs compared to the AC-OPF results. In addition, we show the benefits brought by having four \textit{selected topologies} instead of using only the most recurrent topology identified in the previous topology optimization processes. 

\begin{figure*}[h!]
  \centering
  \begin{minipage}{0.49\textwidth}
    \centering
    \includegraphics[width=\linewidth]{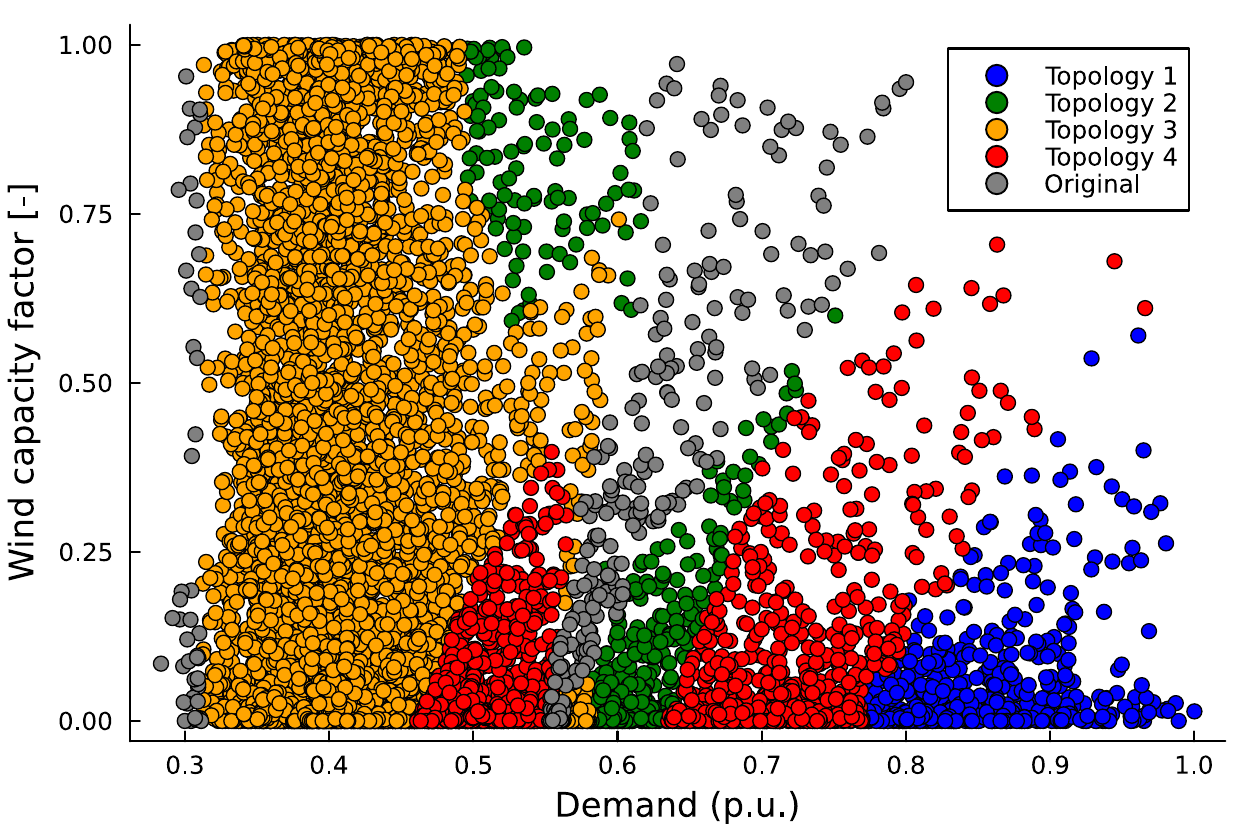}
  \end{minipage}
  \hfill
  \begin{minipage}{0.49\textwidth}
    \centering
    \includegraphics[width=\linewidth]{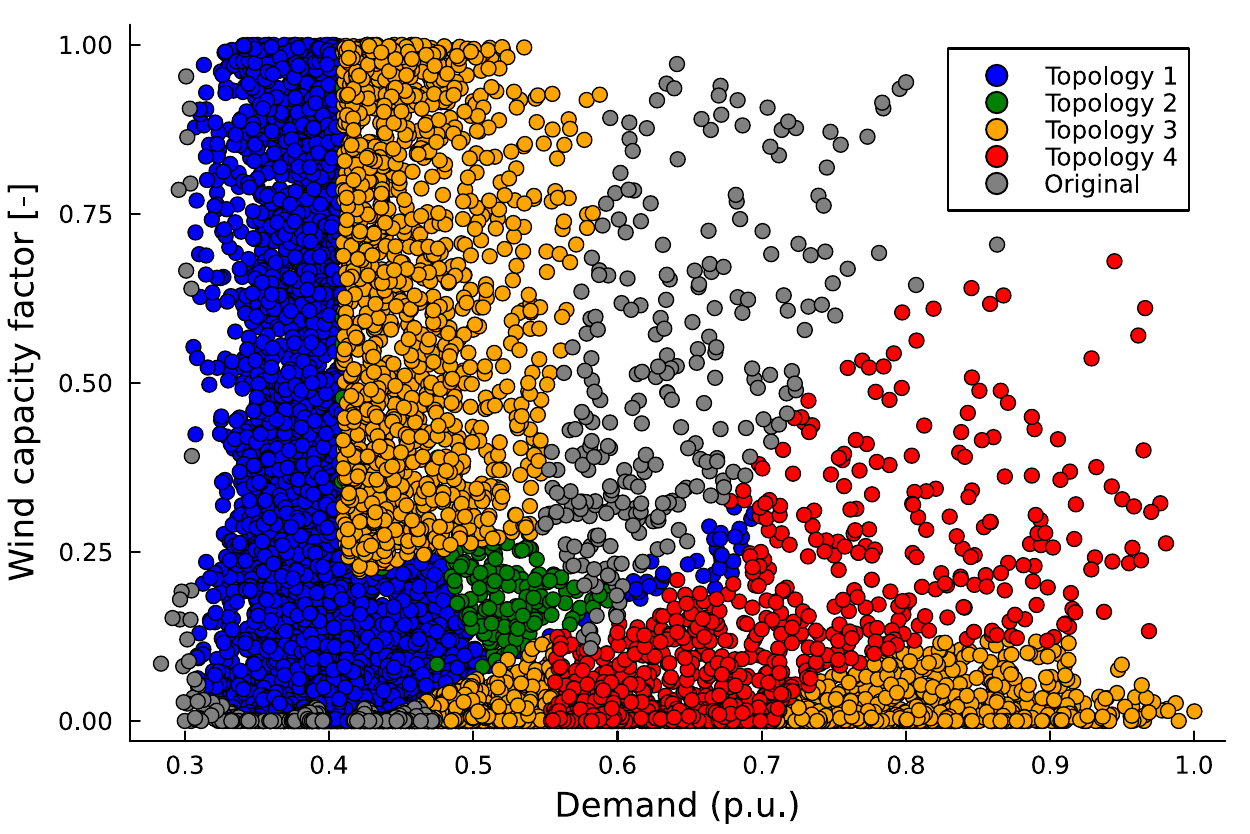}
  \end{minipage}
\caption{\small Optimal busbars \textit{46-49} (left) and \textit{69-24} (right) \textit{selected topologies} with the lowest generation costs in an AC-OPF simulation for each demand-wind combination in \textit{congested} conditions, for a year-long time series. If the \textit{Original} test case topology has the lowest generation costs, the timestep is indicated in gray. Combinations of demand-wind in which different topologies are the most beneficial from a total costs perspective are indicated by the different color distributions along the graph.}
\label{fig:yearly}
\end{figure*}

\begin{figure}[h!]
\centering
\includegraphics[width=0.7\linewidth]{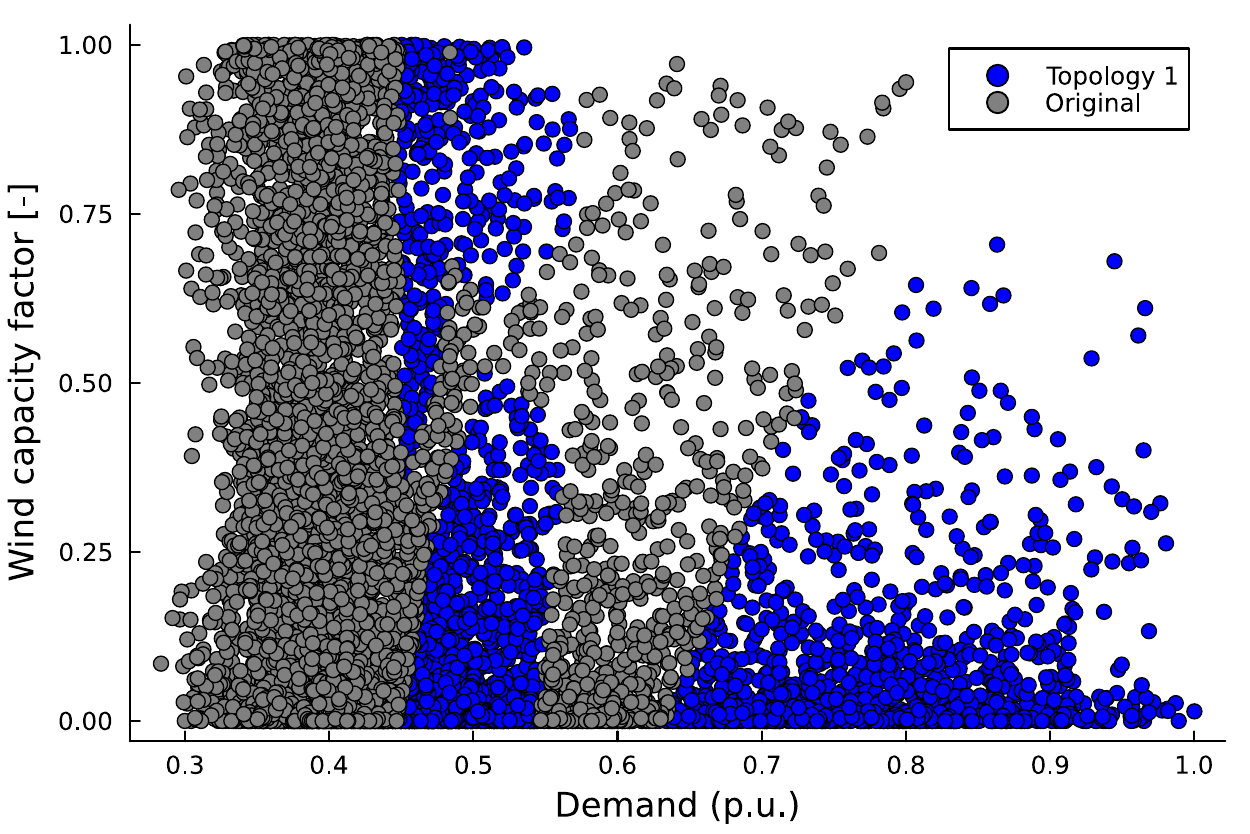}
\caption{\small Optimal topology for every timestep between the most recurrent topology out of the subset of \textit{selected topologies} (\textit{topology 1}) and \textit{original} topologies for the \textit{congested} test case with splittable busbars \textit{46-49}.}
\label{fig:lg_49_46_yearly}
\end{figure}

Firstly, we confirm how the distinction in optimal topologies shown in previous Figures~\ref{fig:46_49_congested} and~\ref{fig:69_24_congested} holds for the year-long time series in Fig.~\ref{fig:yearly} and in Table~\ref{tab:results_optimization_year}. Namely, the wind-demand areas covered by the subset of four \textit{selected topologies} identified for the clustered time series are the same as the ones in the year-long time series. These areas are clearly distinguishable in Fig.~\ref{fig:yearly} thanks to the mapping created by timesteps of different colors.
Moreover, Fig.~\ref{fig:lg_49_46_yearly} shows the most cost-efficient topology for the \textit{congested} test case with splittable busbars \textit{46-49} when only the most recurrent \textit{topology 1} out of the subset of \textit{selected topologies} can be used against the original topology. 

Compared to the left-hand side of Fig.~\ref{fig:yearly},  Fig.~\ref{fig:lg_49_46_yearly} shows that \textit{topology 1} is the most cost-efficient for a larger number of timesteps compared to the case with four \textit{selected topologies} (2856 versus 552 timesteps), but the overall cost reduction is 0.046\% versus 0.117\%, as indicated in Table~\ref{tab:results_optimization_year}. Therefore, allowing a set of four \textit{selected topologies} to choose from leads to a considerable cost reduction compared to using only the most recurrent one in the subset of \textit{selected topologies}. The results in Table~\ref{tab:results_optimization_year} show how this trend is reflected in all the cases we consider in this paper except for \textit{standard, load growth} with splittable busbars \textit{46-49}, where there is no apparent economic value for topology optimization through BuS.
\begin{table*}[h!]
\caption{\small The topology of busbars \textit{46-49} and \textit{24-69} is optimized for four cases, namely \textit{Congested}, \textit{Standard}, \textit{Congested load growth (LG)}, and \textit{Standard load growth (LG)}. Four \textit{selected topologies} ($C_1$ $\rightarrow$ \textit{topology 1}, $C_2$ $\rightarrow$ \textit{topology 2}, $C_3$ $\rightarrow$ \textit{topology 3}, and $C_4$ $\rightarrow$ \textit{topology 4}) are used from the previous clustering of 365 timesteps. The number of timesteps ``\# Timesteps'' indicates in how many timesteps out of 8760 each \textit{selected topology} (including the original one, OPF) leads to the lowest generation costs when running an AC-OPF. The ``Total generation costs'' and ``Curtailment'' are computed for each case for the OPF and topology optimization (TO) models. Their difference is expressed in terms of percentage ``Reduction''.}
\begin{center}
\centering
\resizebox{\textwidth}{!}{%
\begin{tabular}{|c|l|c|c|c|c|c|c|c|c|c|c|c|c|c|c|}
\hline
\multirow{2}{*}{Split} & \multirow{3}{*}{Case} & \multicolumn{5}{c|}{\# Timesteps [h]} & \multicolumn{3}{c|}{Total generation costs $\cdot10^{8}$ [\$]} & \multirow{2}{*}{Simulation} \\
\cline{3-10} 
 \multirow{2}{*}{Buses}& &  \multirow{2}{*}{C1} & 
 \multirow{2}{*}{C2} & 
 \multirow{2}{*}{C3} & 
 \multirow{2}{*}{C4} & 
 \multirow{2}{*}{OPF} &
 \multirow{2}{*}{OPF} &
 \multirow{2}{*}{TO} &
 Reduction &  \multirow{2}{*}{time [h]} \\
 & & & & & & & & & [\%] & \\
\hline
\multirow{8}{*}{46-49} & Congested, 1 topology    & 2856 & - & - & - & 5928 & 984.134 & 981.614 & 0.046 & 0.34 \\
      & Congested, 4 topologies  & 552 & 468 & 5879 & 1590 & 295 & 984.134 & 980.915 & 0.117 & 0.87 \\
      \cline{2-11} 
      & Standard, 1 topology    & 498 & - & - & - & 8286 & 360.318 & 360.315 & 0.001 & 0.37 \\
      & Standard, 4 topologies     & 54 & 1323 & - & - & 7407 & 360.318 & 360.289 & 0.008 & 1.42 \\
      \cline{2-11}  
      & Congested - LG, 1 topology    & 1712 & - & - & - & 7072 & 1048.863 & 1048.013 & 0.081 & 0.38 \\
      & Congested - LG, 4 topologies     & 1492 & 27 & 297 & 3888 & 3080 & 1048.863 & 1047.749 & 0.106 & 1.52 \\
      \cline{2-11}  
      & Standard - LG, 1 topology    & 380 & - & - & - & 8404 & 402.125 & 402.125 & 10$^{-4}$ & 0.38 \\
      & Standard - LG, 4 topologies  & 362 & 134 & - & - & 8264 & 402.125 & 402.125 & 10$^{-4}$ & 1.71 \\
      \hline
    \multirow{8}{*}{24-69} & Congested, 1 topology   & 4447 & - & - & - & 4337 & 984.134 & 983.665 & 0.048 & 0.57 \\
      & Congested, 4 topologies & 4081 & 175 & 2075 & 1389 & 1064 & 984.134 & 983.261 & 0.089 & 1.43 \\
      \cline{2-11} 
      & Standard, 1 topology    & 8777 & - & - & - & 7 & 360.318 & 360.116 & 0.056 & 0.61 \\
      & Standard, 4 topologies  & 8242 & 521 & 2 & 2 & 7 & 360.318 & 360.089 & 0.064 & 1.54 \\
      \cline{2-11}  
      & Congested - LG, 1 topology       & 3823 & - & - & - & 4961 & 1048.863 & 1047.442 & 0.135 & 0.63 \\
      & Congested - LG, 4 topologies     & 2519 & 2190 & 229 & 1103 & 2743 & 1048.863 & 1047.325 & 0.147 & 0.717  \\
      \cline{2-11} 
      & Standard - LG, 1 topology       & 68 & - & - & - & 8692 & 402.125  & 402.101 & 0.006 & 0.60 \\
      & Standard - LG, 4 topologies     & 31 & 77 & - & - & 8686 & 402.125  & 401.170 & 0.008 & 2.60 \\
\hline
\end{tabular}
}
\label{tab:results_optimization_year}
\end{center}
\end{table*}
Finally, Table~\ref{tab:results_optimization_year} includes the number of timesteps in which the \textit{selected topologies} or the \textit{original} topology (\textit{OPF}) are the ones with the lowest generation costs for each of the four investigated cases for the selected couples of split buses \textit{46-49} and \textit{24-69}. Comparing these results to Table~\ref{tab:results_optimization}, the relative amounts of optimal timesteps out of the total for the \textit{4 topologies} in this 8760-hour simulation are consistent with the ones from the clustered time series. In addition, the \textit{1 topology} cases show how selecting only the most recurrent topology out of the subset of \textit{selected topologies} to alternate with the original \textit{OPF} one is still beneficial for a considerable number of timesteps in almost all the investigated cases. This fact results in a reduction in generation costs between 0.046\% and 0.135\% across the \textit{1 topology} cases except for the \textit{Standard}, \textit{46-49} and \textit{Standard - LG}, \textit{46-49}, where the value for grid topology optimization is low due to the loading conditions. When considering \textit{4 topologies}, the generation costs reduction increases to a range of 0.064\% to 0.147\%.

In terms of computational time, selecting among four topologies takes less than two hours for each of the \textit{4 topologies} cases, with all the \textit{1 topology} cases taking less than one hour. Simulations optimizing the topology in real time for every timestep are not included in the results, as actively modifying the topology with different configurations for every timestep is impractical for system operators, and the simulations took more than two days for each case, with only a slight reduction of the total generation costs.

These results confirm how selecting a subset of optimal topologies to choose from can lead to reductions in total generation costs compared to the original OPF simulation, while avoiding long computational time. In addition, from a system operator's perspective, the selected subset of topologies (to be validated with dynamic and N-1 analyses) offers flexibility in operation without the need to compute the optimal topology for each hour, therefore considerably reducing the inherent operational risk related to modifying the grid topology.

\subsection{Topology optimization to increase the busbar hosting capacity with discrete load growth capacity}
In this part of the Section, we investigate whether optimizing the grid topology increases certain buses' \textit{hosting capacity} as the load connected to a certain busbar increases. We assess this by computing the discrete load growth at which load starts being shed for three buses with low LMP in different parts of the test case (\textit{25}, \textit{61} and \textit{100}). We simulate the load growth by increasing the magnitude of the load in steps of 0.2~pu, and compare the results with the original topology. To run the analysis, we select a representative timestep from Fig.~\ref{fig:time_series} with high demand and low wind (timestep \textit{65}), simulating congested conditions in the system as the RES contribution is low. The results for three different simulations, namely OPF with the original topology (\textit{Original}), OPF using one topology out of the subsets of \textit{selected topologies} identified previously (\textit{Pre-selected topology}), and optimizing the grid topology with Model 1 (\textit{Topology optimization}), are shown in Table~\ref{tab:results_load_growth}. Note that the selected configurations for the \textit{Pre-selected topology} simulations are \textit{topology 1} for split buses \textit{46-49} and \textit{topology 4} for split buses \textit{24-69}, as they are the most cost-effective ones for timestep \textit{65} in the two simulations.

\begin{table*}[h!]
\caption{\small Maximum load growth before load shedding at different buses in the 118-bus test case under three simulations: optimal power flow (OPF) with the original topology (\textit{Original}), OPF using one topology out of the subsets of \textit{selected topologies} identified previously in the text (\textit{Pre-selected topology}), and optimizing the grid topology with the proposed topology optimization model (\textit{Topology optimization}). The load shedding point is expressed as a multiple of the maximum single load in the system, connected to bus \textit{116}. The additional hosting capacity is indicated in percentage between brackets. The three gray cells correspond to the ``assigned''  busbar couples for each bus with load growth, identified through the metrics defined in~\cite{Bastianel_2025_PSCC}.
}
\centering
\resizebox{\textwidth}{!}{%
\begin{tabular}{|c|c|c|c|c|c|c|c|c|}
\hline
  & \multicolumn{8}{c|}{\multirow{2}{*}{Point of load shedding}} \\
Bus with & \multicolumn{8}{c|}{} \\
\cline{2-9}
 load growth & & \multicolumn{2}{c|}{Pre-selected topology} & \multicolumn{5}{c|}{Topology optimization -- BuS} \\
\cline{3-9}
  & \multirow{-2}{*}{Original} & 46-49 & 24-69 & 46-49 & 24-69 & 24-49 & 25-66 & 77-80 \\
\hline
  25   & 2.10 [-] & 2.10 (+0.0\%) & 2.10 (+0.0\%) & 2.17 (+3.33\%) & 2.17 (+3.33\%) & \cellcolor{gray_light} 2.24 (+6.67\%) & 2.17 (+3.33\%) & 2.10 (+0.0\%) \\
  61   & 3.39 [-] & 3.39 (+0.0\%) & 3.39 (+0.0\%) & 3.39 (+0.0\%) & 3.39 (+0.0\%) & 3.39 (+0.0\%) & \cellcolor{gray_light} 3.54 (+4.42\%) & 3.47 (+2.36\%) \\
  100  & 3.47 [-] & 3.47 (+0.0\%) & 3.47 (+0.0\%) & 3.47 (+0.0\%) & 3.47 (+0.0\%) & 3.47 (+0.0\%) & 3.47 (+0.0\%) & \cellcolor{gray_light} 3.61 (+4.03\%) \\
\hline
\end{tabular}}
\label{tab:results_load_growth}
\end{table*}

Table~\ref{tab:results_load_growth} indicates that both the \textit{Pre-selected topology} and \textit{Topology optimization} for split buses \textit{46-49} and \textit{24-69} do not bring any benefit compared to the original topology \textit{Original} in terms of \textit{hosting capacity} for each of the three selected buses \textit{25}-\textit{61}-\textit{100}, except for \textit{Topology optimization} in bus \textit{25}. Regarding the three ``assigned'' busbar couples through the metrics proposed in ~\cite{Bastianel_2025_PSCC}, they effectively shift the load shedding threshold by 6.7\% (\textit{24-49} for bus \textit{25}) and 4--4.4\% (\textit{25-66} for bus \textit{61}, \textit{77-80} for bus \textit{100}), as indicated in the diagonal highlighted in gray in Table~\ref{tab:results_load_growth}. For the ``non-assigned'' combinations, only \textit{25--66} for bus \textit{25} and \textit{77--80} for bus \textit{61} increase the load shedding threshold compared to \textit{Original} and \textit{Pre-selected topology} thanks to their proximity to where the discrete load growth takes place.

These results imply that even if selecting an optimal topology can lead to lower generation costs for a certain grid condition, the optimal topology tends to have the same \textit{hosting capacity} as the original topology for several buses over the test case. 

Moreover, the results show that optimizing the topology of busbars selected according to system conditions at the onset of load shedding can increase the \textit{hosting capacity} by several percentage points for a discrete load growth.
Therefore, system operators working on the topic could include the load growth directly in the optimization problem to compute the optimal topology, or use the proposed methodology as a starting point to validate the effective \textit{hosting capacity} for discrete load growth for each bus and related relevant busbars for topology optimization. 
In addition, topological changes could be utilized as transitional measures in case of investment delays for new transmission capacity projects to avoid demand shedding and guarantee a more efficient utilization of the power grid.
\section{Conclusion and future work} \label{sec:conclusion}
This paper presented a methodology to identify a reduced subset of optimal busbar configurations for selected busbars to reduce generation costs while limiting the computational overhead associated with optimizing the grid topology for every timestep.
We applied the proposed approach to two pairs of busbars, \textit{46--49} and \textit{24--69}, in the IEEE 118-bus test case, using wind and demand time series derived from the RTS-GMLC dataset. For each pair, up to four most recurrent \textit{selected topologies} were identified by solving the LPAC-based topology optimization model (plus AC-feasibility checks of the optimized topologies) over 365 representative timesteps, under standard and congested conditions, with and without discrete load growth at bus \textit{69}. Our main findings can be summarized as follows. 

First, a small subset of four busbar \textit{selected topologies} captures significant generation cost savings, achieving cost reductions of up to 0.147\% and curtailment reductions of up to 0.295\% for a set of clustered timesteps compared to a plain AC-OPF simulation with a fixed grid topology. 

Second, the optimal topology is strongly dependent on the wind-demand operating condition, with distinct regions of the demand-wind space consistently favoring specific \textbf{selected topologies}. This pattern persists in the year-long validation time series, confirming the transferability of the identified subsets, with total generation costs reduction up to 0.147\%. 

Third, the value of topology optimization scales with grid utilization: the highest benefits are under congested conditions, and are further amplified by load growth events such as the connection of a large constant load. Under standard, less congested conditions, the original topology remains optimal for the majority of timesteps for the specific test case we studied, limiting the benefit of switching. Local characteristics, such as binding voltage angle constraints, can however, make topology optimization valuable even under standard loading conditions, as observed for busbars \textit{24-69}.

Fourth, the effect of topology optimization on the buses' \textit{demand-growth hosting capacity} for discrete load growth is amplified when the busbars to be optimized are selected based on the system conditions when load shedding starts taking place. Generally, proximity to the load growth influences the relevance of a busbar, too. 

From a practical perspective, the identified subset of topologies, once validated through N-1 contingency and dynamic analyses, offers system operators a manageable set of validated topological actions that can be applied in operations to minimize the generation costs in the system. This considerably reduces the operational risk and decision complexity associated with topological actions, while still delivering considerable economic benefits in terms of generation cost reductions.

Future work will focus on incorporating load growth directly into the topology optimization model to better characterize hosting capacity, and extending the methodology to larger systems with more splittable busbars to achieve more significant reductions in total generation costs, and a more efficient methodology to build a subset of \textit{selected topologies}.


\section*{Acknowledgement}
This paper has received support from the Belgian Energy Transition Fund, FOD Economy, project DIRECTIONS, and from the Research Foundation – Flanders (Belgium) (FWO) travel grant V410025N for Giacomo Bastianel's research visit at the University of Wisconsin-Madison.

\bibliographystyle{ieeetr}
\bibliography{references}

\end{document}